\documentclass[aps,prl,twocolumn,superscriptaddress,floatfix,longbibliography]{revtex4-2}

\usepackage{amsmath,amssymb,mathtools,bm}
\usepackage{graphicx}
\usepackage{booktabs}
\usepackage{float}
\usepackage{xcolor}
\usepackage[colorlinks=true,allcolors=blue]{hyperref}
\usepackage{comment}
\graphicspath{{figs/}}
\allowdisplaybreaks

\newcommand{\dd}{\mathrm{d}}

\newcommand{\avg}[1]{\left\langle #1 \right\rangle}

\newcommand{\Mcal}{\mathsf M}

\begin{document}

\title{Optimal Finite-Time Control of Nonreciprocal Brownian Dimers: Thermodynamic Anomaly and Multiple Transitions}%Optimal finite-time control of nonreciprocal Brownian dimers by two harmonic traps}

\author{Ruicheng Bao}
\email{ruicheng@g.ecc.u-tokyo.ac.jp}
\affiliation{Department of Physics, Graduate School of Science,
The University of Tokyo, Hongo, Bunkyo-ku, Tokyo 113-0033, Japan}

\begin{abstract}
We solve exactly a finite-time thermodynamic optimal control problem for two nonreciprocally interacting Brownian particles translated by two %independently moving
harmonic traps.  The controller manipulates both the center and separation of the pair. Nonreciprocal interactions generate an internal active force that couples these two channels. The optimal protocol is oscillatory, deliberately opens the dimer even when the target separation is unchanged, and can extract work during transport. A central finding is a finite critical time beyond which the external-work infimum is $-\infty$: at any prescribed duration beyond this threshold, both extractable work and output power are unbounded. %This anomaly arises from competition between the nonreciprocity and viscous drag.  
Physical regularizations such as finite trap range and force saturation restore a finite optimum and convert the anomaly into
optimal-protocol transitions: in the zero-target case, a hard finite range produces a first-order-like jump from the zero protocol to a maximum-range protocol, whereas smooth force saturation gives a continuous, second-order-like onset. Under finite-range constraints, the optimal protocol can further undergo
multiple finite-time transitions, producing multiple work-duration kinks with no qualitative analog in prior studies. %Under finite-range constraints, the optimal protocol can further undergo multiple finite- time transitions,  reflected in more than one kink in the optimal work-duration curve.

 %We also derive work fluctuations and extend the conclusions to general interactions, higher dimensions, and three or more harmonic traps.
\end{abstract}

\maketitle

\noindent\textit{Introduction---}At mesoscopic scales, energetic cost and operating speed are central performance measures in biological,
chemical, and physical processes. Stochastic thermodynamics \cite{Seifert2012,peliti2021stochastic,cao2025Stochastic} provides a versatile framework for
quantifying the tradeoff between these two figures of
merit in mesoscopic systems. Building on this framework, finite-time thermodynamic optimal control \cite{SchmiedlSeifert2007,11prl_optimal,sivak23optimal} offers a quantitative
language for designing fast yet energetically efficient
manipulations of mesoscopic objects, including diverse
biomolecules and colloidal assemblies \cite{Bustamante2005,Seifert2012,Ciliberto2017}.

Most results on the finite-time thermodynamic optimal control concern one controlled particle \cite{SchmiedlSeifert2007,gomez2008optimal,schmiedl2008efficiency,11prl_optimal,OlsenLowen2025,GarciaMillan2025ClosedLoop,25pre_loos,loos_26transitions,24prx_loos,25pnas_loos,pietzonka26accuracy}. For a single passive particle in a moving harmonic trap, the optimal protocol has symmetric endpoint jumps and a linear interior \cite{SchmiedlSeifert2007,gomez2008optimal}. Recent nonequilibrium extensions show that a prescribed time-dependent force \cite{OlsenLowen2025}, activity \cite{GarciaMillan2025ClosedLoop,25pre_loos} or spatial heterogeneity \cite{loos_26transitions} can be incorporated exactly. Further extensions to systems with memory effects were given in \cite{24prx_loos,25pnas_loos}. %A complementary route is provided by optimal transport theory, when the controller can prescribe the full probability distribution rather than only a finite set of trap parameters \cite{sivak23optimal}. 
For interacting systems, only a few results exist, mostly based on perturbative methods \cite{PhysRevX.14.011012,26arxiv_loos}. By contrast, exact finite-time control beyond a single body remains much
less developed, despite its importance for realistic systems and the quantitative control of active biological matter.%exact finite-time control beyond a single particle remains much less developed, even though it is a key step toward the understanding of optimal finite-time processes in realistic systems and the quantitative control of active biological matter.%even though it is a key step toward thermodynamically optimal finite-time processes in realistic interacting systems and toward the quantitative understanding and control of active biological matter. %The difficulty is also the opportunity: internal coordinates can be actuated and can feed back into transport, so interacting systems need not reduce to independent single-particle controls.

\begin{figure}[H]
 \centering

\includegraphics[width=0.6\linewidth]{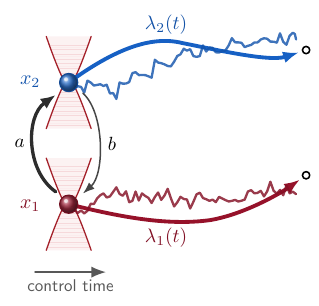}
 \caption{Two overdamped Brownian particles are held by independently movable harmonic traps. The interaction is reciprocal when $a=b$, and nonreciprocal when $a\ne b$.}
 \label{fig1}
\end{figure}

Here we take a step forward by exactly solving a minimal interacting model. Two overdamped Brownian particles, coupled by either a reciprocal or nonreciprocal interaction, are confined by two harmonic traps whose centers can be independently moved from prescribed initial to prescribed final positions (see Fig. \ref{fig1}). Nonreciprocal interactions can arise naturally in nonequilibrium systems with sustained energy input, including chemical, active-matter, and biological systems \cite{Ivlev2015,Loos_2020,Fruchart2021,23prr_nonreciprocal,dechant25ncomm}. We seek the driving protocols that minimize the mean work. The setup is motivated by the dual-trap optical tweezers ubiquitous in single-molecule experiments \cite{94nature_dual,shaevitz03_nature,moffitt2008recent,ribezzi14free,bustamante2021optical,23ncomm_dual}, where simultaneous control of two traps is allowed to realize translation or separation of the trapped pair. %but we use the more general language of harmonic traps because the theory depends only on harmonic confinement.

For reciprocal interactions, center and extension control channels separate, reducing the problem to one-body transport plus interaction free-energy storage. In contrast, nonreciprocity couples the channels through an active area term, yielding oscillatory protocols and asymmetric endpoint jumps.  Beyond a critical duration, this term drives the external-work infimum to $-\infty$, so both extractable work and mean
output power are unbounded without regularization. Finite trap range or force saturation restores
boundedness and yields first/second-order-like transitions. For finite-range controls, finite duration can select distinct optimal
strategies, yielding repeated protocol transitions unique to interacting systems.

\noindent\textit{Model---}We consider a one-dimensional overdamped dimer in
dimensionless units, with trap stiffness, friction, and bath temperature
set to unity. Dimensional restoration is given in Supplemental Material
(SM) \cite{supplemental_material} Sec.~S1. The particle positions obey
\begin{subequations}\label{eq:micro_model}
\begin{align}
 \dot x_1&=-(x_1-\lambda_1)+a(x_2-x_1-q_0)+\sqrt{2}\,\xi_1(t),\\
 \dot x_2&=-(x_2-\lambda_2)-b(x_2-x_1-q_0)+\sqrt{2}\,\xi_2(t),
\end{align}
\end{subequations}
where $\xi_{1}(t)$ and $\xi_2(t)$ are independent Gaussian white noise processes and $q_0$ denotes both the initial trap separation and the interaction rest length. %about which the interaction is linearized. 
The reciprocal case is $a=b$.  We define $\rho=1+a+b>0,\  
\alpha=\frac{a-b}{2}$. For fixed trap centers, the dynamics is stable whenever
$\rho=1+a+b>0$. In what follows, we restrict to the extension-restoring
regime $\rho\ge 1$. 

Let
\begin{equation}
 Y=\frac{\lambda_1(t)+\lambda_2(t)}{2},
 \quad
 R(t)=\lambda_2(t)-\lambda_1(t)-q_0,
\end{equation}
and shift the initial center to zero. The target center displacement and target excess separation are
\begin{equation}
 Y(0)=0,
 \quad R(0)=0,
 \quad
 Y(T)=L,
 \quad R(T)=M .
\end{equation}
%\rev{Unlike a single particle, an interacting dimer generally relaxes after the protocol to particle means offset from the final trap centers when $M\ne0$; for the linear model the offset vanishes for rigid translation, $M=0$, as detailed in SM Sec.~S11.}
The mean center and mean excess extension are
\begin{equation}
 y=\avg{\frac{x_1+x_2}{2}},
 \quad
 s=\avg{x_2-x_1}-q_0 .
\end{equation}
Averaging Eq.~\eqref{eq:micro_model} gives the deterministic dynamics
\begin{subequations}\label{eq:mean_dyn}
\begin{align}
 \dot y&=-(y-Y)+\alpha s,\label{eq:y_dyn}\\
 \dot s&=-\rho s+R .\label{eq:s_dyn}
\end{align}
\end{subequations}
\noindent\textit{Objective: the mean work---}The mean external work is
\begin{equation}\label{eq:mean_work}
 W[Y,R]=\int_0^T\left[2\dot Y(Y-y)+\frac12\dot R(R-s)\right]\dd t .
\end{equation}
This is a rewriting of the original definition of the external work  $W:=\sum_{i=1}^2\int_0^{T}\dot\lambda_i(\lambda_i-\avg{x_i})dt$.

The controls can be eliminated in favor of the mean path, using Eq.~\eqref{eq:mean_dyn}.
The trap endpoints are fixed, but $Y$ and $R$ may jump at $t=0$ and $t=T$, whereas $y$ and $s$ are continuous.  %Carrying the jump work explicitly and integrating the smooth interior by parts gives the analytical expression of the mean work. 
For any piecewise-smooth two-trap protocol satisfying $Y(0)=R(0)=0$ and $Y(T)=L$, $R(T)=M$, the mean work equals
\begin{widetext}
\begin{equation}\label{eq:path_functional}
 W[y,s]=\int_0^T \left(2\dot y^2-2\alpha s\dot y+\frac12\dot s^2\right)\dd t
 +(L-y_T)^2+\frac14(M-s_T)^2+\frac{\rho-1}{4}s_T^2,
\end{equation}
\end{widetext}
where $y(0)=s(0)=0$, $y_T=y(T)$, and $s_T=s(T)$. General initial mean positions are treated in SM \cite{supplemental_material} Sec.~S2. Notably, the term $-2\alpha\int_0^{T}s\dot{y}dt$ is the only nonreciprocal contribution. It is the active channel by which extension can help or oppose center transport. A minimizer $(y^*,s^*)$ of Eq.~\eqref{eq:path_functional} yields the optimal interior controls by substitution into a rewritten form of Eq.~\eqref{eq:mean_dyn}:
\begin{equation}\label{eq:controls_from_path}
 Y^*=y^*+\dot y^*-\alpha s^*,
 \quad
 R^*=\dot s^*+\rho s^* .
\end{equation}

\noindent\textit{Reciprocal dimers---}When $\alpha=0$, the two variational channels separate.  The optimal interior paths and controls are
\begin{subequations}\label{eq:rec_protocols}
\begin{align}
 y^*(t)&=\frac{Lt}{T+2},
 &Y^*(t)&=\frac{L(1+t)}{T+2},\label{eq:rec_center}\\
 s^*(t)&=\frac{Mt}{\rho T+2},
 &R^*(t)&=\frac{M(1+\rho t)}{\rho T+2},\label{eq:rec_relative}
\end{align}
\end{subequations}
for $0<t<T$, with symmetric initial and final jumps to the prescribed endpoints.  The minimum work is
\begin{equation}\label{eq:W_rec}
 W_{\rm rec}^*=\frac{2L^2}{T+2}
 +\frac{M^2[2+(\rho-1)T]}{4(\rho T+2)} .
\end{equation}
The first term is exactly two copies of the one-particle result \cite{SchmiedlSeifert2007}.  The second term is the finite-time cost of changing the trap separation against a reciprocal linker.

The minimum work to independently control two free particles is
\begin{equation}\label{eq:W_free}
 W_{\rm free}^*
 =\frac{2L^2+M^2/2}{T+2} .
\end{equation}
Thus, a reciprocal dimer is equivalent to two free particles for rigid translation, $M=0$. In general,
\begin{equation}\label{eq:rec_excess}
 W_{\rm rec}^*-W_{\rm free}^*
 =\frac{M^2T^2(\rho-1)}{4(T+2)(\rho T+2)} .
\end{equation}
Recall that $\rho\ge1$, so the difference is nonnegative and the reciprocal linker cannot reduce the work relative to two independently trapped particles.  %If $0<\rho<1$, the reciprocal interaction is locally softening and the same formula is negative: changing the trap separation then releases interaction free energy and can save work, but this regime requires the external traps to stabilize the relative coordinate.
In the quasistatic limit the free-particle contribution vanishes, whereas the reciprocal dimer retains
\begin{equation}\label{eq:DeltaF_rec}
 \Delta F_{\rm rec}=\lim_{T\to\infty}W_{\rm rec}^*
 =\frac{M^2(\rho-1)}{4\rho},
\end{equation}
the equilibrium free energy stored in the interaction.

\begin{figure}[H]
 \centering
\includegraphics[width=0.9\columnwidth]{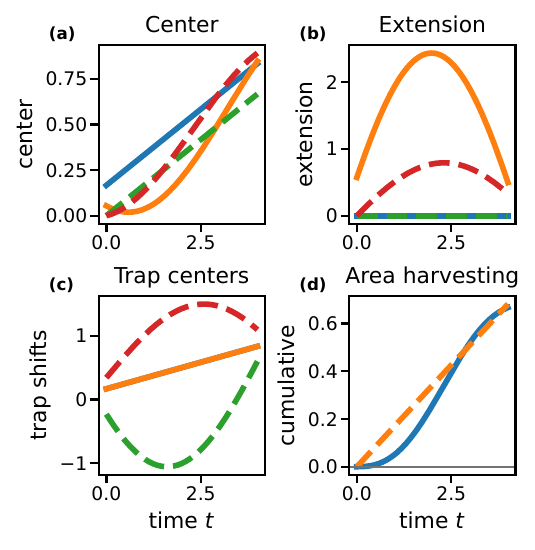}
 \caption{Rigid target translation with $L=1$, $M=0$, $T=4$, and $\rho=3$.  In panels (a,b), solid curves are trap controls and dashed curves are mean particle coordinates; blue/green denote the reciprocal dimer and orange/red the nonreciprocal dimer.  Panel (c) shows the shifted individual trap centers: solid curves are reciprocal and dashed curves are nonreciprocal.  The nonreciprocal dimer with $\alpha=0.6$ deliberately opens the extension channel even though the target separation is unchanged. Panel (d) shows the cumulative area-harvesting gain $2\alpha\int_0^t s\dot y\,\dd u$ (blue) against the positive drag part (orange dashed).}
 \label{fig:protocols}
\end{figure}

\noindent\textit{Nonreciprocal dimers: oscillatory optimal paths and protocols---}For $\alpha\ne0$, the Euler-Lagrange equations and terminal natural boundary conditions are
\begin{subequations}\label{eq:EL_main}
\begin{align}
 &\ddot y=\frac{\alpha}{2}\dot s,
 \quad\ddot s=-2\alpha\dot y,\label{eq:EL_bulk}\\
 &2\dot y_T-\alpha s_T-(L-y_T)=0,
 \quad2\dot s_T+\rho s_T-M=0 .\label{eq:EL_bc}
\end{align}
\end{subequations}
Equivalently, the center velocity $v=\dot y$ satisfies
\begin{equation}\label{eq:oscillator_v}
 \ddot v+\alpha^2 v=0 .
\end{equation}
This oscillator is generated by area term $-2\alpha s\dot y$. The controller deliberately excites the extension channel during transport to harvest the nonreciprocal force.%The controller rotates the pair $(\dot y,s)$ in the center-extension plane to harvest the nonreciprocal force.

For $0<t<T$, the optimal mean paths for the controlled particles read 
\begin{subequations}\label{eq:nonrec_solution}
\begin{align}
 y^*(t)&=\frac{A\sin(\alpha t)+B[1-\cos(\alpha t)]}{\alpha},\label{eq:y_sol}\\
 s^*(t)&=\frac{2\{A[\cos(\alpha t)-1]+B\sin(\alpha t)\}}{\alpha},\label{eq:s_sol}
\end{align}
\end{subequations}
Consequently, one has $\dot y^*(t)=A\cos(\alpha t)+B\sin(\alpha t)$ and $\dot s^*(t)=2[-A\sin(\alpha t)+B\cos(\alpha t)]$. Let $\theta=\alpha T$, $C=\cos\theta$, and $S=\sin\theta$. $A$ and $B$ are determined by substituting the optimal solution into Eq.~\eqref{eq:EL_bc}, yielding:
\begin{equation}\label{eq:AB_system}
\underbrace{\begin{pmatrix}
2+S/\alpha & (1-C)/\alpha\\
-4S+2\rho(C-1)/\alpha & 4C+2\rho S/\alpha
\end{pmatrix}}_{\Mcal_\alpha(T,\rho)}
\begin{pmatrix}A\\ B\end{pmatrix}
=\begin{pmatrix}L\\ M\end{pmatrix}.
\end{equation}
The optimal controls $R^*(t)$ and $Y^*(t)$ then follow from applying Eq.~\eqref{eq:controls_from_path}. See Fig.~\ref{fig:protocols}(a)-(c) for an illustration of the optimal mean paths and protocols. 

The constants have physical meaning. From Eq.~\eqref{eq:nonrec_solution}, $A=\dot y^*(0^+)$ and $B=\dot s^*(0^+)/2$, i.e., $A$ is the initial center-trap jump and $B$ is half the initial extension-trap jump, $Y^*(0^+)=A$ and $R^*(0^+)=2B$.  The final jumps are fixed by the same constants but with a rotation determined by the nonreciprocity $\alpha$, so the initial and final jumps are asymmetric for nonreciprocal interactions \footnote{It is useful to view the jumps in the two collective control channels as a vector. Define the initial and final jumps as $\Delta_0Y=Y(0^+)-Y(0),\quad \Delta_0R=R(0^+)-R(0)$ and $\Delta_TY=Y(T)-Y(T^-),\quad \Delta_TR=R(T)-R(T^{-})$, respectively. If
\begin{equation*}
 \bm j_0=\begin{pmatrix}\Delta_0Y\\ \Delta_0R/2\end{pmatrix}=\begin{pmatrix}A\\B\end{pmatrix},
 \qquad
 \bm j_T=\begin{pmatrix}\Delta_TY\\ \Delta_TR/2\end{pmatrix},
\end{equation*}
then the optimality equations give \begin{equation*}
    \bm j_T=
 \begin{pmatrix}\cos(\alpha T)&\sin(\alpha T)\\-\sin(\alpha T)&\cos(\alpha T)\end{pmatrix}\bm j_0.
\end{equation*} Thus the single-particle and reciprocal symmetry of equal initial and final jumps is recovered at $\alpha=0$, whereas a nonreciprocal dimer generically has unequal component-wise endpoint jumps: the jump vector keeps its norm in the $(Y,R/2)$ plane but is rotated during the protocol.  The shifted individual trap jumps $\Delta\tilde\lambda_{1,2}=\Delta Y\mp\Delta R/2$ inherit this asymmetry. See SM Sec. S5 for more details.}. 

The optimal work has the explicit form (See End Matter)
\begin{subequations}
\begin{align}
 &W_{\rm nr}^*=L[L-y^*(T)]+\frac{M}{4}[M-s^*(T)]\label{eq:W_nonrec1}\\ &=\frac{K (4L^2+M^2)+(\rho-1)QM^2-4\alpha QLM}{\alpha^2\det\Mcal_\alpha}\label{eq:W_nonrec},
\end{align}    
\end{subequations}
with $Q\equiv1-C,\ K\equiv2\alpha^2C+\alpha\rho S$. Here, the first $K$ term is the finite-time transport cost renormalized by the nonreciprocity. The second term is the terminal free-energy storage in the interaction. The last cross term is purely nonreciprocal and is proportional to the signed area of the rectangle spanned by the center and extension displacement.  

\begin{figure}[H]
\centering \includegraphics[width=0.8\columnwidth]{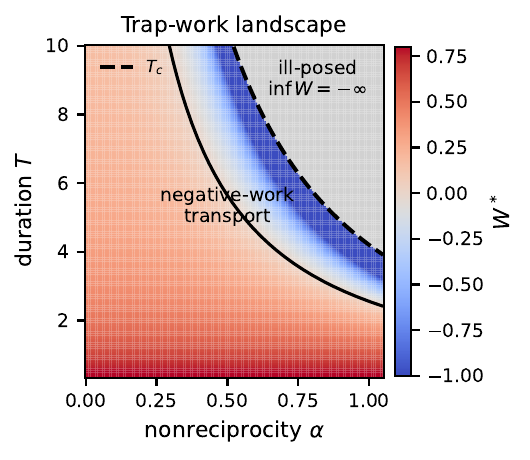}
 \caption{Minimum trap-work landscape for $\rho=3$, $L=1$, and $M=0$.  The solid contour marks $W^*=0$.  The dashed curve is $T_c$.  To its right the unregularized trap-work problem has no finite minimum.} %physical constraints such as those in Fig.~\ref{fig:regularization} must be included.}
 \label{fig:landscape}
\end{figure}

\noindent\textit{Thermodynamic anomaly and its restoration---}The determinant of Eq.~\eqref{eq:AB_system} is
\begin{align}\label{eq:det_main}
 &\det\Mcal_\alpha=\frac{4}{\alpha^2}\Delta_\alpha(T,\rho),\\
 &\Delta_\alpha(T,\rho)\equiv2\alpha^2\cos(\alpha T)\nonumber
 \\&+\alpha(\rho+1)\sin(\alpha T)+\rho[1-\cos(\alpha T)]\nonumber.
\end{align}
Let $T_c$ be the first positive zero of $\Delta_\alpha$ \footnote{Throughout, we refer to \(T_c\) as the critical duration to emphasize
that the total protocol duration is prescribed. In optimal-control
terminology, it is the first conjugate time.}.  For $T<T_c$, the second variation is positive definite for all admissible
perturbations. Since the work functional is quadratic, Eq.~\eqref{eq:nonrec_solution} is the unique global minimizer. At
$T=T_c$, this quadratic form develops a zero direction. In general, the initial and final jumps (namely, $A$ and $B$) diverge as $(T_c-T)^{-1}$ when
$T\to T_c^-$ (see SM Sec. S6). Thus the infimum of the external work is unbounded below \footnote{For
nongeneric endpoints, for example the zero-target case $L=M=0$, the prescribed target vector $(L,M)$ can have zero projection onto the singular direction at $T=T_c$. Then the jumps need not diverge exactly at $T=T_c$; the quadratic form is only semidefinite and the minimizer is degenerate. For any $T>T_c$, however, the
present linear problem still has a negative-curvature direction and is unbounded below.}. For $T>T_c$, the second variation has a negative direction, so the infimum remains $-\infty$.  Since $\lim_{\alpha\to 0}\det\Mcal_\alpha>0$ and $\lim_{\alpha\to 0}W^*_{\rm nr}=W^*_{\rm rec}\geq0$, there is no finite $T_c$ for reciprocal cases. See Fig. \ref{fig:landscape}. 

The stable unregularized branch always ends before one velocity period, $T_c<2\pi/|\alpha|$, see SM Sec. S6.  Stronger coupling, $|\alpha|\ge\sqrt{\rho}$, further gives $T_c\le\pi/|\alpha|$, forbidding any peak-trough pairs; for weaker coupling the stable window can extend beyond half a period, and even one individual trap center can contain one peak-trough pair for suitable endpoints.

The physical origin of the anomaly is the competition, illustrated in
Fig.~\ref{fig:protocols}(d), between drag dissipation and the
nonreciprocal contribution $-2\alpha\int_0^T s\dot y\,\dd t$ in
Eq.~\eqref{eq:path_functional}.  When $T>T_c$, the protocol is long
enough that a suitable excursion can harvest this active contribution more
efficiently than it pays drag; increasing the excursion amplitude then
amplifies this imbalance without bound. For
$T<T_c$, the same area term may still reduce the work, and even produce
finite negative work, but any amplitude increase is countered at least as
fast by the drag cost, so the unbounded mechanism is absent (\cite{supplemental_material} Sec. S6).

%The physical origin of the anomaly is as follows. A closed or nearly closed loop harvests trap work through the oriented area $-2\alpha\oint s\,\dd y$ in the $(y,s)$ plane [c.f. Eq.~\eqref{eq:path_functional}]. For long protocols the dissipative terms can be made small compared with this area term, and scaling the loop amplitude then drives $W$ to $-\infty$.  The extracted trap work is not free energy from the visible dimer; it is work drawn from the hidden nonequilibrium mechanism that maintains nonreciprocity. For $T<T_c$, however, every area loop still has positive quadratic cost: a large extension must be opened and closed too quickly, so the positive drag penalties dominate the nonreciprocal area reward.

\begin{figure}[H]
\centering \includegraphics[width=0.9\columnwidth]{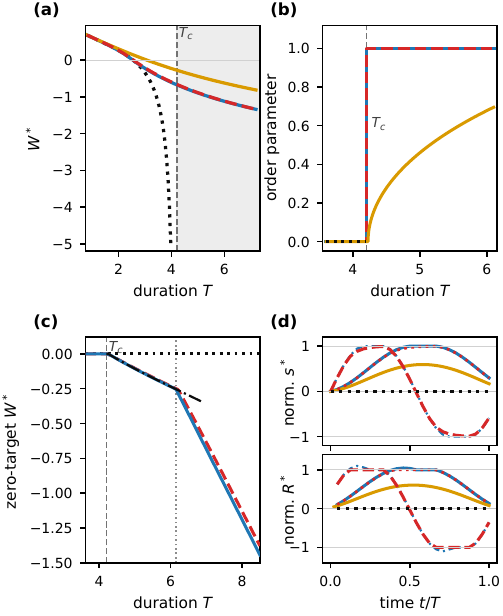}
 \caption{Physical regularizations of the $T>T_c$ anomaly for
\(\alpha=1\) and \(\rho=3\). Blue denotes a hard cutoff, yellow a smooth saturating force, and red a finite-range constraint. (a) For \(L=1,M=0\), the
regularized works remain finite beyond \(T_c\). Black dotted curve denotes the
unregularized curve. (b) For $L=M=0$ (zero-target), finite-range constraints show a step-like first-order onset, while smooth
saturation turns on continuously. Order parameters are $\|R^*\|_\infty/R_{\max}$ or $\|s^*\|_\infty/s_{\max}$.  (c) Zero-target work for
finite-range regularizations. Gray dotted line marks the switch time from the one-wall to two-wall strategy, black dotted horizontal line the zero protocol, and dash-dotted line the one-wall protocol.
(d) Representative optimal protocols (normalized by $R_{\max}$, $s_{\max}$ or $s_0$).  Black dotted zero protocol, $T=3.80$; blue solid hard-cutoff one-wall branch, yellow smooth-saturation branch, and red dotted finite-range one-wall branch, $T=5.55$; blue dash-dotted hard-cutoff two-wall branch and red dashed finite-range two-wall branch, $T=7.35$.
}
 \label{fig:regularization}
\end{figure}

Two physical regularizations are natural.  First, real traps can move the separation only over a finite range.  We therefore impose the experimentally relevant constraint $|R(t)|\le R_{\max}$, so the stable equation $\dot s=-\rho s+R$ implies $|s(t)|\le R_{\max}/\rho$ when $s(0)=0$. Then $2\dot y^2-2\alpha s\dot y\ge-\alpha^2s^2/2$ yields the lower bound \footnote{the contribution of the boundary term is non-negative for $\rho\geq 1$.}
\begin{equation}\label{eq:range_bound}
 W\ge -\frac{\alpha^2R_{\max}^2}{2\rho^2}T,
\end{equation}
and the work anomaly is removed. A hard cutoff $|s|\le s_{\rm max}$ gives a similar mechanism with $s_{\rm max}$ replacing the reachable scale $R_{\max}/\rho$; it is included in Fig.~\ref{fig:regularization} as a direct comparison.  Second, the active force from nonreciprocity may saturate in the real world, as in molecular motors. Currently, the mean active force is $\alpha s$, which diverges as $s$ diverges. Replacing that with a saturating function of $s$, for instance, $g_{\rm sat}(s)=\alpha s_0\tanh(s/s_0)$, can also restore the finite optimum. Incorporating the energy input that maintains nonreciprocity into the objective function is another possible regularization. Since the active force is typically treated as a resource to be harnessed \cite{GarciaMillan2025ClosedLoop,OlsenLowen2025}, this regularization is only detailed in SM Sec. S7.

Notably, the results above extend qualitatively to higher spatial dimensions, and systems with more particles
and more traps (SM Sec.~S8).

\noindent\textit{First- and second-order-like protocol transitions and multiple kinks---}After regularization, the competition between nonreciprocity
and drag can turn the anomaly into sharp transitions in the optimal protocol. The cleanest case is the zero-target cycle, $L=M=0$, where the zero
protocol is a stationary solution whose stability changes at $T_c$. For a hard range, the quadratic form is positive for $T<T_c$, so the zero protocol is the global optimum.  For $T>T_c$, a negative-curvature mode can be amplified until a wall is reached, producing a first-order-like jump of the normalized amplitude, e.g. $\|R^*\|_\infty/R_{\max}$ or $\|s^*\|_\infty/s_{\max}$, from $0$ to $1$, as shown in Fig. \ref{fig:regularization}(b). 

For smooth saturation, the order parameter instead turns on continuously but with a singular first derivative at $T_c$,
\begin{equation}\label{eq:smooth_landau_scaling_main}
 \|s^*\|_\infty/s_0\propto (T-T_c)^{1/2},
\end{equation}
characteristic of a second-order-like transition (SM Sec. S9).

Remarkably, the protocol can have multiple transitions in the duration, as reflected by two or more kinks in the work-duration curves. For example, Fig.~\ref{fig:regularization}(c) shows a second kink, where the optimal protocol topology switches from one-wall (the protocol only saturates $R_{\rm max}$) to two-wall (the protocol saturates both $\pm R_{\rm max}$). The curve can even have more than two kinks in higher spatial dimensions or
in systems with more particles and more traps; nonzero targets ($L>0$) also exhibit protocol transitions (SM Sec.~S9).

Finally, defining $P^*_{\rm ext}(T)=-W_{\rm ext}^*(T)/T$, the output power approaches a long-time plateau set by the maximum available nonreciprocal force $\lim_{T\to \infty}P^*_{\rm ext}=\alpha^2R_{\max}^2/(2\rho^2)$. Interestingly, finite-time local shoulders or peaks can appear due to the transient competition between area harvesting and drag. See SM Sec.~S7.

\noindent\textit{Discussion---}Nonreciprocal two-body optimal control reveals a range
of phenomena absent from one-body or equilibrium control.
Nonreciprocity produces oscillatory optimal protocols that take a longer route and exhibit asymmetric endpoint jumps. Above a
critical duration, the optimal work becomes unbounded below,
giving a novel thermodynamic anomaly. This anomaly is a
diagnostic: it marks where the coarse-grained nonreciprocal force
must be physically regularized. Finite trap range or force saturation
restores a finite optimum and converts the singularity into sharp protocol transitions. Under finite-range constraints, the optimum can switch repeatedly as the duration varies,
producing multiple finite-time transitions and corresponding
work-duration kinks. This is a genuinely interacting phenomenon without counterpart in single-particle control.%Reciprocal interactions are invisible to pure center transport, but not to extension control. Nonreciprocity turns extension into a controlled active force on the center.  Without physical regularizations, the original optimization problem can become ill posed, namely, the optimal work can be unbounded above a critical duration.  The singularity is a diagnostic: it identifies the regime where a coarse-grained nonreciprocal model must be modified by incorporating appropriate regularizations such as finite-trap range and force saturation.

Experimentally, dual-trap actomyosin platforms \cite{94nature_dual,18ncomm_dual,23ncomm_dual} provide a natural route to test our theory: the two optical traps can control the filament translation and separation, while the myosin motor supplies a nonequilibrium drive that may be described as an effective nonreciprocal interaction at the coarse-grained level. Recent nonreciprocally interacting nanoparticle experimental platforms provide additional complementary candidates \cite{rieser22_Science,livska24nphy,reisenbauer_24nphy}.

Finally, searching for thermodynamic anomalies and multiple protocol transitions in discrete systems \cite{Barato_2017,deweese_21pre,dechant2022minimum,jann2026near,dechant2026renormalized}, quantum systems \cite{18pra_quantum,Rolandi_2023}, and other types of optimal control problems \cite{11prl_optimal,sivak23optimal,23prx_tan,deweese_24prl,23prl_ABPcontrol,pietzonka26accuracy} is an interesting direction. We expect such phenomena to arise when two or more competing mechanisms
favor distinct optimal strategies.

\begin{acknowledgments}
R. B. was supported by JSPS KAKENHI Grant No. 25KJ0766.  R. B. is grateful to Qixiao Yuan for checking the derivations. %Some results were derived with assistance from ChatGPT.
\end{acknowledgments}

\bibliography{refs}

\begin{center}{\large\bfseries{End Matter}}\end{center}

\noindent\textit{Explicit constants and work decomposition---}For $T<T_c$, $\Delta_\alpha>0$ and Eq.~\eqref{eq:AB_system} can be inverted explicitly.  With $Q=1-C$, one obtains
\begin{subequations}\label{eq:AB_explicit_end}
\begin{align}
 A&=\frac{\alpha\{(4\alpha C+2\rho S)L-QM\}}{4\Delta_\alpha},\\
 B&=\frac{\alpha\{[4\alpha S+2\rho Q]L+(2\alpha+S)M\}}{4\Delta_\alpha} .
\end{align}
\end{subequations}
These constants are the initial center velocity and one half of the initial extension velocity, respectively.  The terminal means are
\begin{subequations}\label{eq:terminal_explicit_end}
\begin{align}
 y_T^*&=\frac{(\alpha S+\rho Q)L+(\alpha Q/2)M}{\Delta_\alpha},\\
 s_T^*&=\frac{2\alpha QL+(Q+\alpha S)M}{\Delta_\alpha} .
\end{align}
\end{subequations}
Substitution into Eq.~\eqref{eq:W_nonrec1} gives the closed quadratic form, Eq.~\eqref{eq:W_nonrec}. This formula separates the passive finite-time transport coefficient $K/\Delta_\alpha$, the reciprocal free-energy storage proportional to $(\rho-1)QM^2$, and the nonreciprocal target-geometry interference $-\alpha QLM$.  For rigid transport, $M=0$,
\begin{equation}\label{eq:W_M0_end}
 W_{\rm nr}^*(M=0)=L^2\frac{2\alpha^2 C+\alpha\rho S}{\Delta_\alpha}\le \frac{2L^2}{T+2},
\end{equation}
and the inequality strengthens monotonically with $|\alpha|$ on the stable branch. When $M=0$ the admissible set is symmetric under $s\to -s$, so minimizing the work functional $W_0[y,s]-2|\alpha|\,|\int s\dot y\,\dd t|$ can only lower the value as $|\alpha|$ grows ($W_0$ is the work with $\alpha=0$, see SM Sec. S5 for details).  For $M\ne0$ the odd cross term in Eq.~\eqref{eq:W_nonrec} records whether the imposed extension change helps or fights the nonreciprocal transport direction.

\end{document}